\documentclass[smallextended]{svjour3}
\smartqed
\usepackage[dvips]{graphicx}
\usepackage[usenames,dvipsnames]{color}
\usepackage{epsfig}
\usepackage{rotating}
\usepackage{amssymb}
\usepackage{amsmath,amsfonts}
\usepackage[boxed]{algorithm}
\usepackage{verbatim}
\usepackage{booktabs}

\newtheorem{thm}{Theorem}
\newtheorem{prop}{Proposition}

\newtheorem{cor}{Corollary}

\newcommand{\Z}{\mathbb{Z}}

\newcommand{\RR}{\mathbb{R}}
\newcommand{\pf}{{\noindent \bf Proof. \ }}

\title{Constructive Spherical Codes\\
near the Shannon Bound}
\author{Patrick Sol\'{e}\and Jean-Claude Belfiore}
\institute{ Patrick Sol\'{e}\at Telecom ParisTech, CNRS LTCI\\
Communications and Electronics Dept.\\
46 rue Barrault, 75634 Paris CEDEX 13, France\\{\em Present address:} MECAA,\\Math Dept of King Abdulaziz University,\\Jeddah, Saudi Arabia\\
\email{sole@enst.fr} 
\and Jean-Claude Belfiore \at Telecom ParisTech, CNRS LTCI\\
Communications and Electronics Dept.\\
46 rue Barrault, 75634 Paris CEDEX 13, France\\ \email{belfiore@enst.fr} 
}

\begin{document}

\maketitle

\begin{abstract}
Shannon gave a lower bound in 1959 on the binary rate of spherical codes of given minimum Euclidean distance $\rho$.
Using nonconstructive codes over  a finite alphabet, we give a lower bound that is weaker but very close for small values of $\rho.$
The construction is based on the Yaglom map combined with
some finite sphere packings obtained from nonconstructive codes for the Euclidean metric.
 Concatenating geometric codes meeting the TVZ bound
with a Lee metric BCH code over $GF(p),$ we obtain spherical codes that are polynomial time constructible. Their parameters outperform those obtained by Lachaud and Stern 
in 1994. At very high rate they are above $98$ per cent of the Shannon bound.
\end{abstract}
\section{Introduction}
A {\bf spherical code} is a finite set of points of the unit sphere in a Euclidean space of finite dimension. For motivation and background see \cite{EZ,SPLAG}.
Let $X$ denote a spherical code  of ${\RR}^n$. Denote by $\rho$ its Euclidean squared minimum distance. Define its binary rate
as $$R(\rho):=\limsup \frac{\log_2(\vert X\vert)}{n}. $$ Chabauty in 1953 and Shannon in 1959 \cite{LS}, gave a lower bound 

$$R(\rho)\ge R_S(\rho):=1-(1/2)\log_2(\rho (4-\rho)).$$

Lachaud and Stern, in 1994 gave a lower bound on the rate $R_*(\rho)$  of {\em polynomial time constructible} spherical codes as

$$R_*(\rho)\ge 0.5 R_S(\rho).$$

In the present, work we shall give a lower bound based on nonconstructive lattice packings

$$R(\rho)\ge R_L(\rho):=-(1/2)\log_2(\rho ).$$

It can be shown by direct substitution that $$R_S(\rho)\ge R_L(\rho)$$ for all $0\le \rho \le 4.$

However, for small $\rho$, the two curves are very close to each other:

$$R(\rho)- R_L(\rho)=O(\rho^2).$$

We give a family of spherical codes based on non constructive codes for the Euclidean metric whose asymptotic performance is just as good.

$$R(\rho)\ge R_L(\rho),$$

This shows that using finite alphabet codes for constructing finite packings is as efficient as using truncation of dense infinite sphere packings.

However, the main result of this work is a lower bound based on an explicit family of polynomial time constructible codes for the Euclidean metric that outperforms 
the Lachaud Stern bound almost by a factor of $2$
$$R_*(\rho)\ge 0.98 R_S(\rho).$$

for some very small values of $\rho$ in the range $0<\rho \le 1.$ These codes, furthermore admit efficient encoding and decoding algorithms.

The article is organized as follows. In the next section we explore the Yaglom map and use it on truncations of infinite lattices. In Section III we study the performance 
of long codes for the Euclidean metric over a fixed alphabet. The analysis technique is based on the saddle point method. In Section IV we give the explicit construction of long
such codes and prove it is better not only than the relevant Gilbert bound but also after Yaglom map than $98$ per cent of the Shannon bound.
\section{The Yaglom map}
Following Yaglom \cite[Chap. 9, Thm. 6]{SPLAG}, we inject ${\RR}^n$ into ${\RR}^{n+1}$ by the map

$$Y:\; x \mapsto (x, \sqrt{R^2-x\cdot x}).$$
Note that this map sends the ball $B(n,R)$ of  radius $R$ that is 
$$B(n,R):=\{ x \in {\RR}^n \vert \, x\cdot x\le R^2\} $$
in ${\RR}^n$ into the sphere $S(n,R)$ of radius
 $R$ in ${\RR}^{n+1}$
$$ S(n,R):=\{ x\in {\RR}^{n+1} \vert \, x\cdot x= R^2\}$$
 and that distance between points can only increase. The following result is then immediate.
{\prop \label{fonda}
If $P$ is a packing of spheres of diameter $d$ in $B(n,R)$ then $Y(P)$ is a spherical code in $S(n,R)$ of minimum square distance $d^2.$}

Using lattice packings we can prove our first bound.

{\thm There are families of spherical codes built from lattice packings by the Yaglom map such that $$R(\rho)\ge R_L(\rho):=-(1/2)\log_2(\rho ).$$}

\pf
By the Yaglom bound \cite[Chap. 9, Thm. 6]{SPLAG} we know there are spherical codes satisfying
$$R\ge \delta+1+R_L(\rho),$$
where $\delta$ is the asymptotic exponent of the best density of lattice 
packings in ${\RR}^n$, or $\frac{\log_2(\Delta)}{n}$ in the notation of \cite[Chap. 1]{SPLAG}. Now by the Minkowski 
bound \cite[Chap. 1, (46)]{SPLAG},
 we know that $\delta \ge -1.$ The result follows.
\qed

Using lattice packings built from codes over fields by the so-called Construction A \cite[p.182]{SPLAG}  we obtain our second bound.
{\thm There are families of spherical codes built from Construction A lattice packings such that $$R(\rho)\ge R_L(\rho).$$}

\pf

By \cite{Ru} we know there are lattice packings built from codes over prime fields by Construction A having asymptotic density

$$ \delta \ge -1.$$ 

 The result follows, as the preceding one, upon applying the Yaglom bound.
\qed

This result will be obtained also for finite sphere packings made of codes for the Euclidean metric (Theorem (\ref{bigalpha})).
Using {\bf explicit} lattice packings constructed by AG techniques  we can prove our last bound. Note that, even though these lattices are polynomial time constructible,
the resulting spherical codes may not be because of the cost of truncation which may require an exponential time algorithm like Finke-Pohst \cite{FP}.

{\thm There are families of spherical codes of binary rate $R(\rho)$ built from lattice packings by the Yaglom map such that $$R(\rho)\ge R_L(\rho)-1.30.$$}

\pf
By the Yaglom bound \cite[Chap. 9, Thm. 6]{SPLAG} we know there are spherical codes satisfying
$$R\ge \delta+1+R_L(\rho),$$
where $\delta \ge -2.30$ is the asymptotic exponent of the best density of lattice
packings constructed in \cite{LT}.
The result follows.
\qed

{\cor For $\rho \le e^{-90}$ there are spherical codes constructed from polynomial time constructible lattices with $R\ge 0.98 R_S(\rho).$  }

\pf
We apply the preceding Theorem with the approximation $R_S(\rho)\approx R_L(\rho).$ We obtain the said bound for $\rho \le 2^{-130}\approx e^{-90}.$
\qed
\section{Codes for the Euclidean metric}
We consider codes of length $n$ over the integers modulo an integer $q.$ If $q=2s+1$ is odd we represent $\Z_q$ on the real line by the constellation
$$\{-s, \cdots,-1,0,1,\cdots,s.\}$$ If $q=2s+2$ is even we represent $\Z_q$ on the real line by the constellation
$$\{-s-1/2, \cdots,-3/2,-1/2,1/2,\cdots,s-1/2,s+1/2\},$$
which is a shift of the natural representation
$$\{-s, \cdots,-1,0,1,\cdots,s,s+1\}$$ by $-1/2.$
We denote by $\phi $ the induced map from $\Z_q^n$ into $\RR^n.$ If $q$ is odd then $\phi(x)\cdot \phi(x) \le n s^2,$ while if $q$ is even
$\phi(x)\cdot \phi(x) \le n (s+1/2)^2,$ We consider the Euclidean distance $d_E(,)$ on $\Z_q^n$ defined by $d_E(x,y)=(\phi(x)-\phi(y))^2.$
This distance could have been defined as induced by the standard Euclidean weight $w_E()$ on $\Z_q$ that is $w_E(x)=\min(x^2,(q-x)^2.$
{\prop \label{euclid}
If $C$ is a code of length $n$ and minimum Euclidean  distance $d$ over $\Z_q$ then $\phi(C)$ is a packing of spheres of diameter $\sqrt{d}$
of $B(n,s\sqrt{n})$ if $q$ is odd and of $B(n,(s+1/2)\sqrt{n})$ if $q$ is even.}

Combining with Proposition \ref{fonda} we obtain

{\prop \label{useful}
If $C$ is a code of length $n$ and minimum Euclidean  distance $d$ over $\Z_q$ then $Y(\phi(C))$ is a spherical code of squared Euclidean distance $d$
of $S(n,s\sqrt{n})$ if $q$ is odd and of $S(n,(s+1/2)\sqrt{n})$ if $q$ is even.}

Let $V(n,q,r)$ denote the size of the ball of radius $r$ for the Euclidean metric in $\Z_q^n$. The following result is the analogue of the standard Gilbert bound 
in that setting.

{\prop \label{Gilbert}There are codes in $\Z_q^n$ of Euclidean distance $d$ and  of cardinality $$\vert C\vert\ge \frac{q^n}{V(n,q,d-1)}.$$}

The technically difficult part is to estimate the asymptotic exponent of $V(n,q,r)$ for $n$ large and $r=\lfloor \lambda n\rfloor.$ To that end, we introduce 
the following generating series $f(z)=1+2 s(z)$ with $$s(z)=\sum_{i =1}^sz^{i^2}$$ for odd $q$ and $$s(z)=\sum_{i =1}^{s-1}z^{i^2}+z ^{s^2}$$ for even $q.$

{\thm \label{estim} The asymptotic exponent of $V(n,q,r)$ for $n$ large and $r=\lfloor \lambda n\rfloor$ is

$$\lim(\frac{\log_2(V(n,q,r))}{n}) =\log_2(f(\mu))-\lambda\log_2(\mu) $$ where $\mu$ is the unique real positive solution of $zf'(z)=\lambda f(z).$}

\pf
 The generating function for the numbers $V(n,q,r)$ is of the form $f(z)^ng(z)$ with $g(z)=1/(1-z).$ The result follows by application of \cite[Corollary 2]{GS}.
\qed


{\thm \label{main} With the preceding notation, there are spherical codes of the unit sphere $S(n+1,1)$  with relative squared Euclidean distance $\rho\le 1$ of binary rate
$$R\ge \log_2(q)-\log_2(f(\mu))+a\rho\log_2(\mu)$$
 where $\mu$ is the unique real positive solution of $zf'(z)=a\rho f(z),$ with $a=s^2$ for $q$ odd and $a=(s+1/2)^2$ for $q$ even.
}

\pf
 We combine the Gilbert bound of Proposition \ref{Gilbert} with the estimate of Theorem \ref{estim} to construct long Euclidean metric codes with prescribed parameters
and the Yaglom map of Proposition \ref{useful} to derive spherical codes from them. Note that we let $\lambda=a \rho$ to rescale $S(n+1,\sqrt{na})$ to $S(n+1,1).$
\qed

This approach can be generalized to the case of a large varying alphabet.

{\thm \label{bigalpha} There  are spherical codes, constructed from codes over $\Z_q,$ with $q$ odd and variable and asymptotic rate  $$R(\rho)\ge R_L(\rho)-c,$$
with $c\approx 0.77 \times 10^{-8}.$}

\pf
 We use the same saddle point analysis as \cite{RS} on the generating function
 $$1+2\sum_{i =1}^\infty z^{i^2}.$$
Note that $s$ does not occur in the latter expression since $q\ge 2r+1.$
\qed
\section{Constructive bound}
By the TVZ bound \cite[Th. 13.5.4]{HP} we know there are families of geometric codes over $GF(Q)$ for $Q$ a square with rate ${\cal R}$ and relative distance $\Delta$ satisfying

$$ {\cal R}+\Delta \ge 1- \frac{1}{\sqrt{Q}-1}.$$

We concatenate this geometric code with a code over $\Z_q$ of parameters $[n,k]$ and minimum Euclidean distance $d_E.$ 
We must assume therefore that $Q=q^k.$ If $q$ is not a prime, we label $\Z_q$ by the elements of $GF(q)$ in an arbitrary fashion.
In order to apply the TVZ bound we must assume $Q$ to be square. 

{\prop \label{concat} With the above notation, the  Yaglom map of the concatenated code has parameters $(R,\rho)$ above the straight line
$$ \frac{Rn}{\log_2(q)k}+ \frac{\rho ns^2 }{d_E}\ge 1- \frac{1}{\sqrt{q^k}-1}$$}

\pf
We assume some familiarity with concatenation \cite[\S 5.5]{HP}.
The $q-$ary rate of the concatenated code is $\frac{k}{n} {\cal R}.$ The binary rate of the spherical code is therefore
$R=\log_2(q) \frac{k}{n} {\cal R}.$
The relative distance of the inner Euclidean metric code is
 $ \frac{d_E}{n}.$
The relative distance of the concatenated code is $ \Delta \frac{d_E}{n}.$
After normalization to reduce to the unit sphere Proposition (\ref{euclid}) yields
$$\rho \ge \Delta \frac{d_E}{n s^2}.$$
  Substituting into the TVZ bound we are done.
\qed







Let us take, as inner codes, the Roth-Siegel BCH codes \cite{RoS}. The Lee minimum distance is an immediate lower bound on the Euclidean minimum distance of such a code.
{\thm \label{TVZ} For each prime $p\ge 7$ and every integer $1\le t \le (p+1)/2,$ such that $p$ is congruent to $t+1 \mod{2},$ there is a family of spherical codes with binary rate $R$ and minimum squared Euclidean distance $\rho$ over $\Z_p$ satisfying
$$\frac{R(p-1)}{(p-t-1)\log_2(p)}+\frac{\rho (p-1)^3}{8t} \ge 1- \frac{1}{p^{(p-t-1)/2}-1} .$$
}
\pf
 For these codes $n=p-1,$ $d_L\ge 2t$ and $k\ge n-t,$ by \cite{RoS}. 
\qed
We need the following elementary result whose proof is omitted.

{\prop\label{calculus} The equation of the tangent in $\rho$ of the curve $(\rho,\lambda R_L(\rho))$ is of the form $$\frac{X}{A}+ \frac{Y}{B}=1, $$ with
\begin{eqnarray*}
 A&=& \rho (1-\ln( \rho))\\
B&=&\frac{\lambda}{2\ln(2)} (1-\ln( \rho))
\end{eqnarray*}
}
We are now in a position to state and prove the main result of this note.
{\thm Take $p=5432455719452623343140299649993224712642268405087\\
972148236533041723675544652674874508958455203602044198462638584629866\\
4106668659730094751$ and $t/n=0:00155359$ in Theorem (\ref{TVZ}). For all values of $\rho \le e^-{640:48}$ and $\lambda=0.98$ the bound 
of Theorem (\ref{TVZ}) is strictly above the tangent
to the curve $(\rho,\lambda R_L(\rho))$ at $\rho.$ 
}

\pf
Denote by $f=1- \frac{1}{p^{(p-t-1)/2}-1}$ a number very close to $1.$ Put $\tau=t/n.$
We find values of $p$ and $\rho$ that the intersections of the tangent with the axes are less than 
\begin{eqnarray*}
 B&\le & \frac{(p-t-1)\log_2(p)}{p-1}\\
A& \le& \frac{8\tau f}{(p-1)^2}
\end{eqnarray*}

\begin{figure}[ht]

\noindent \begin{centering}
\includegraphics[width=0.85\columnwidth]{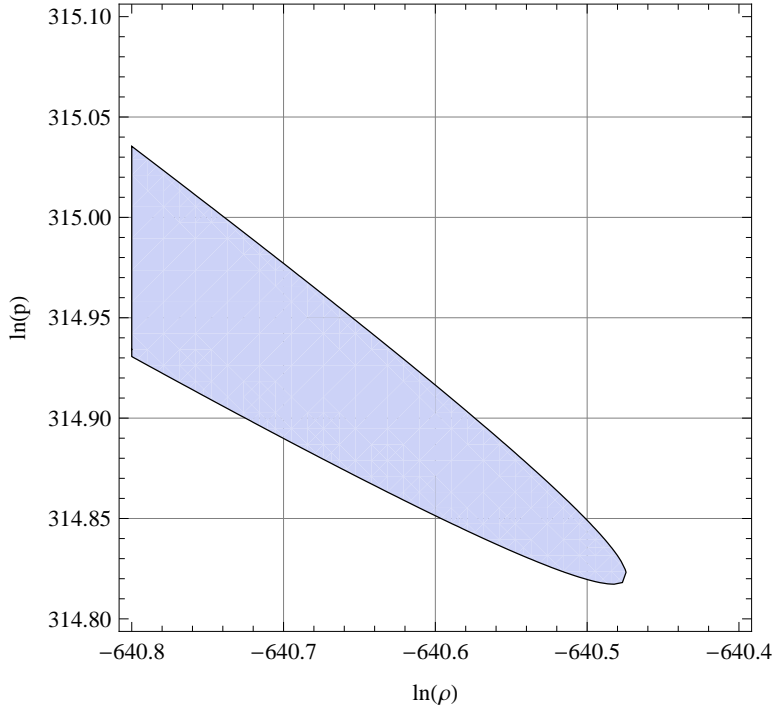}
\par\end{centering}

\caption{\label{fig:Attainable-Region}Attainable Region for $\ln\rho$ and
$\ln p$; here, $\lambda=0.98$}
\end{figure}

Approximating $p-1$ by $p$ and $f$ by $1$ we get by replacing $A,\,B$ by their expressions from Proposition \ref{calculus},
\begin{eqnarray*}
\lambda( 1-\ln(\rho))& \le & 2 (1-\tau)\ln(p)\\
\rho (1-\ln(\rho))& \le& \frac{8\tau }{p^2}
\end{eqnarray*}
Write 
\begin{eqnarray*}
x&= & \ln(\rho)\\
y&=& \ln(p)
\end{eqnarray*}
After this change of variables, and getting rid of $\tau,$ between
$$(1-x)\exp(x+2y) /8\le \tau \le 1-\lambda (1-x)/2y $$
 we obtain the equation
$$\exp(x+2y)(1-x)+4\lambda (1-x)/y\le 8,$$
which can be solved numerically and graphically as shown in Figure \ref{fig:Attainable-Region}. 
Once the values of $x$ and $y$ have been constrained in that way the  assertions of the Theorem are checked by direct (machine) computation.
\qed

Recall that the envelope of a family of curves is a curve all the points of which are tangent at some point to one of the curves in the family.
We can obtain the envelope of the family of straight lines promised by TVZ when $p$ varies as follows

{\thm Some points produced by Theorem \ref{TVZ} lie on the curve
$$\frac{R\ln(2)}{(1-\tau) y}+\frac{1}{1-x}=1$$ where

\begin{eqnarray*}
x&= & \ln(\rho)\\
y&=& \ln(p)\\
x+2y&=& c\\
8\tau &=& (1-x)\exp(c)
\end{eqnarray*}
and $c$ is a constant depending on the range of $\tau.$
}
\pf
Keep the notation of the preceding proof.
An approximate equation for the straight line of Theorem \ref{TVZ} is 
$$\frac{R}{(1-\tau) \ln_2(p}+\frac{p^2}{8 \tau}=1$$
We choose arbitrarily
$$x+2y=c$$ for some constant $c$ and
 $$8\tau = (1-x)\exp(c).$$
The result follows afer some algebra.
\qed
\begin{figure}[ht]

\noindent \begin{centering}
\includegraphics[width=0.85\columnwidth]{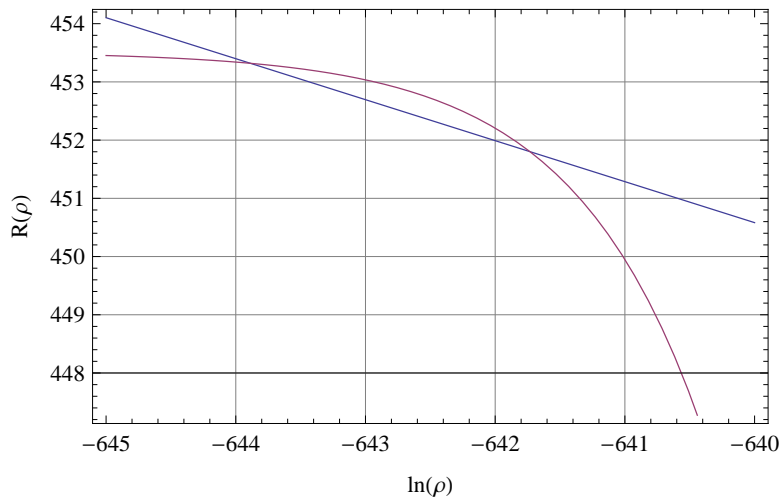}
\par\end{centering}

\caption{\label{fig:R-rho}Rate as a function of $\ln\rho$ for both our constructive codes and $\lambda \times$ the Shannon bound for $\lambda=0.976$}

\end{figure}
Experimentally, the curve obtained in Figure \ref{fig:R-rho} shows that our constructive codes outperform $\lambda \times$ the Shannon bound. Note that we
took, here, a value of $\lambda=0.976$ to show more clearly the performance of such codes, even if these codes have been designed for $\lambda=0.98$.
\section{Complexity issues}
While the size of the large prime in Theorem 8 might seem outlandish, we believe that our constructions of spherical codes might have some practical interest
for more realistic inner codes.
We observe that the complexity of the Yaglom map is $O(N)$ and that the concatenated codes being linear, encoding of the spherical codes is linear in time and
space.
 This contrasts with most standard constructions of spherical codes, especially those based on constant weight codes \cite{EZ}.
As for decoding, the decoding of concatenated code is possible knowing decoding algorithms for the inner and outer codes \cite{D}. As explained 
in \cite[\S 13.5.2]{HP} polynomial time decoding algorithms are known for certain codes obtained from the Garcia Stichtenoth towers of function fields.
An algebraic decoding algorithm up to the Lee error correcting capacity  is provided for the BCH codes we use in \cite{RoS}.
\section{Conclusion}
In this work we have constructed spherical codes by using finite packings of spheres within a ball and mapping them on the surface of
 the sphere in the next dimension by using the Yaglom map. The construction of finite packing was done by using codes for the Euclidean metric.
 We hope this will stimulate research in this area. For instance can the minimum Lee distance bounds
of \cite{RoS} for BCH codes be improved for the Euclidean metric? Are there perfect codes for the Euclidean metric?
This might provide stronger inner codes for the concatenation construction and might lead to
an improvement of the Shannon lower bound.

{\bf Acknowledgement} The authors thank Christine Bachoc and Philippe Gaborit for helpful discussions.

\end{document}